\documentstyle[psfig]{mn}
\newcommand{\reference}{\bibitem}
\def\beq{\begin{equation}}
\def\eeq{\end{equation}}

\def\kpc{\,{\rm {kpc}}}
\def\kpch{\,{h^{-1}{\rm kpc}}}
\def\mpch{\,h^{-1}{\rm {Mpc}}}
\def\kms{\,{\rm {km\, s^{-1}}}}

\def\msun{{\rm M}_\odot}

\def\v200{{V_{200}}}
\def\r200{r_{\rm vir}}
\def\M200{M_{\rm vir}}
\def\s100{{\cal S}_{100}}

\def\e0{{\epsilon_0}}

\def\pc{\rm pc}
\def\msun{\rm M_\odot}
\input{psfig.sty}
\begin{document}
\title[Understanding the Results of Galaxy-Galaxy Lensing]
{Understanding the Results of Galaxy-Galaxy Lensing using
Galaxy-Mass Correlation in Numerical Simulations}
\author[Yang, Mo, Kauffmann, \& Chu]
{XiaoHu Yang$^{1,2,3}$, H. J. Mo$^{1}$, Guinevere Kauffmann$^{1}$,
 YaoQuan Chu$^{2,3}$
\\
\smallskip
      $^1$Max-Planck-Institut f\"ur Astrophysik
      Karl-Schwarzschild-Strasse 1, 85748 Garching, Germany \\
      $^2$Center for Astrophysics, University of Science and Technology
      of China, Hefei, Anhui 230026, China\\
      $^3$National Astronomical Observatories, Chinese Academy of Science,
      Chao-Yang District, Beijing, 100012, China
      }
\date{Accepted ........
      Received .......;
      in original form .......}
\maketitle

\begin{abstract}
McKay et al. (2002) have recently used measurements of weak
galaxy-galaxy lensing in the Sloan Digital Sky Survey to estimate
the cross correlation between galaxies and the projected dark
matter density field. They derive a relation between aperture mass
within a radius of  $260\kpch$, $M_{260}$, and lens galaxy
luminosity, that does not depend on galaxy luminosity, type or
environment. In this paper, we study the cross-correlation between
galaxies and dark matter using galaxy catalogs constructed from a
high-resolution N-body simulation of a $\Lambda$CDM Universe. We
show that our simulations reproduce the McKay et al. results
reasonably well. In the simulation, $M_{260}$ is approximately
equal to the halo virial mass for $L_*$ galaxies. $M_{260}$ 
overestimates the virial mass for fainter galaxies and underestimates
it for brighter galaxies. We use the simulations to show that
under certain circumstances the halo virial mass may be recovered
by fitting an NFW profile to the projected galaxy-mass correlation
function. If we apply our method to the observations, we find that
$L_*$ galaxies typically reside in halos of  $\sim 2\times 10^{12}
h^{-1} \msun$, consistent with halo masses estimated from the
observed Tully-Fisher relation. In the simulations, the halo
virial mass scales with galaxy luminosity as $L^{1.5}$ for central
galaxies in halos and for galaxies in low-density regions. For
satellite galaxies, and for galaxies in high-density regions,
there is no simple relation between galaxy luminosity and halo
mass and care must be exercised when interpreting the lensing
results.
\end{abstract}

\section{Introduction}

In a recent study, McKay et al. (2002, herafter M2002) used the
signal of galaxy-galaxy lensing detected in the Sloan Digital Sky
Survey (SDSS) to infer the dark matter mass distribution around
bright galaxies. What McKay et al. measured from the data was the
shear field $\gamma_+$, inferred from the distorted images of
background galaxies around lensing galaxies. From the theory of
weak lensing (e.g. Schneider et al., 1999), the shear field is
related to the surface mass density contrast $\Delta\Sigma_+$ as:
\begin{equation}
\gamma_{+}(R)\Sigma_{crit} = \overline{\Sigma}(\leq R)
    - \overline{\Sigma}(R) \equiv \Delta\Sigma_+\,,
\end{equation}
where $\overline{\Sigma}(\leq R)$ is the mean surface density
within radius $R$, $\overline{\Sigma}(R)$ is the 
mean surface density at radius $R$,
and $\Sigma _{\rm crit}$ is the critical density
determined by the geometry of the lens-source system. Defined this way,
$\Delta\Sigma_+$ is a measure of the cross correlation between the
lens galaxies and the projected mass density. We will therefore
refer $\Delta\Sigma_+$ as the galaxy-mass correlation function
(GMCF).

The GMCF depends on the matter distribution around lens galaxies.
The observed GMCF can therefore be used to infer the extent of
dark haloes around galaxies (e.g. Hudson et al. 1998; Fischer et
al. 2000; Smith et al. 2001; Wilson et al. 2001; Guzik \& Seljak
2002). Using the SDSS data, McKay et al. (M2002) found that: (i)
the shear field $\gamma_+$ could be measured with reasonable
signal-to-noise out
    to a projected radius $\sim 1 \mpch$;
(ii) $\Delta\Sigma_+(R)$ could be fit by a power law,
    with the singular isothermal sphere an acceptable model for
    the distribution of dark matter around galaxies;
(iii) the mass $M_{260}$ (defined as the mass 
    projected within an aperture of radius $260\kpch$)
    inferred from the observed GMCF was strongly correlated with the
    luminosity of lens galaxies and in the redder photometric pass-bands,
    the galaxy luminosity was proportional to $M_{260}$, with a
    proportionality constant of order 100;
(iv) the ratio of aperture mass to luminosity in the red photometric bands did not
    depend significantly on galaxy luminosity, type and environment.

Because  current models of galaxy formation based on the cold dark
matter (CDM) cosmogony predict the existence of extended dark
haloes around galaxies, the results of M2002 provide important
constraints on models of galaxy formation.

The observed GMCF cannot be interpreted simply
in terms of individual halo profiles. There are three other factors that can
affect it. First, the observed GMCF is an average of
mass density fields around different galaxies.
Second, the GMCF depends
on how lens galaxies are distributed within dark haloes. Third,
more than one halo can contribute to the projected density field
around a given galaxy.
These effects must be quantified in order
to use the observed GMCF to infer the properties of galaxy haloes.

In this paper, we use a galaxy catalogue constructed from a
high-resolution N-body simulation to understand the observed
galaxy-galaxy lensing results. We first examine whether or not the
simulation results agree with the observations. We then illustrate
how the three effects mentioned above influence the GMCF. Finally,
we discuss what kind of information about dark haloes it is
possible to infer from the GMCF. The paper is organized as
follows. In Section 2 we describe the simulation data.
Calculations of the GMCFs for various samples are presented in
Section 3. In Section 4, we investigate the relation between the
masses derived from the GMCFs and the actual masses of dark haloes
in the simulation. We then use these relations to estimate the
halo masses of observed lens galaxies. Finally, in Section 5,  we
summarize and discuss our results.

\section{The Simulations}

In this paper, we use the results of the GIF simulation of the
$\Lambda$CDM model [with
$(\Omega_0,\Lambda,\Gamma,\sigma_8) = (0.3,0.7,0.21,0.90)$] carried
out by Kauffmann et al. (1999) using codes from
the Virgo Consortium (Jenkins et al., 1998; 
the data are publicly available
\footnote{http://www.mpa-garching.mpg.de/Virgo/}.) The
simulation evolves $256^3$ cold dark matter particles in a
periodic cube of side length 141.3h$^{-1}$Mpc, with particle mass
$1.4\times 10^{10} h^{-1}M_{\odot}$, and with a force softening
$\eta=20\kpch$. Dark haloes were selected from the
simulation using a ``friends-of-friends'' (FOF) algorithm with a
linking length of $0.2$ times the mean inter-particle separation,
and galaxy catalogues were constructed by populating dark haloes
with galaxies according to a semi-analytic model of galaxy
formation described in Kauffmann et al. (1999a). Model galaxies are
assigned luminosities and morphological types according to their
star formation histories and bulge-to-disk ratios.
Previous investigations with these simulated catalogues
demonstrated that the model is quite successful in matching the
observed properties of the galaxy population, both at the
present day and at high redshifts (Kauffmann et al.,
1999a, b; Diaferio et al., 1999; 2001).

For this study, we use new publically available galaxy
catalogues constructed from the GIF simulations
giving predictions for the SDSS photometric bands.
We present results only for the $z$-band. Calculations were also made for the
$r$ and $i$-bands, but the results are very similar. Fig. 1 shows
the luminosity distribution of the model galaxies. Results are
shown for `central galaxies' (which are the brightest galaxies
located at the centers of dark matter halos) and for `all
galaxies' (which include both central galaxies and satellites). As
can be seen, the bright end of the luminosity function is
dominated by central galaxies, while satellite galaxies dominate
the faint end. Because of the mass resolution of the simulation,
predictions for galaxies fainter than $-19.5$ mag are uncertain.
Such galaxies are excluded from our analyses.
\begin{figure}
\centering \vskip-0.5cm
\psfig{figure=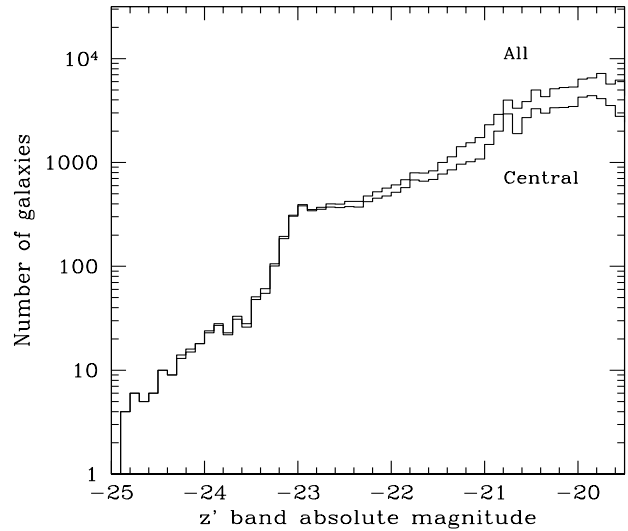,width=8.5cm,height=8.5cm,angle=0}
\vskip-0.5cm \caption{The distribution of GIF galaxies with
respect to the $z$-band absolute magnitudes.
Results are shown separately for central galaxies and
all galaxies.}
\end{figure}

\section{Analyses}

\subsection{The GMCF}

As explained in Section 1, the observed shear fields around
galaxies can be used to derive the cross-correlation between lens
galaxies and the projected mass density field.
In this paper, we calculate the GMCF directly from the mass and
galaxy distributions in the simulations.
We thus neglect all the uncertainties involved in reconstructing
the mass-density field from the weak lensing observations.

We first project the positions of galaxies and dark matter
particles onto a plane (chosen to be one surface of the simulation
box). For each galaxy, we estimate the mean surface density
contrast of dark matter within rings of different radii around the
galaxy. The GMCF as a function of radius, $\Delta\Sigma_+ (R)$, is
obtained by averaging over a given  sample of galaxies. Since by
definition the background surface density is subtracted from
$\Delta\Sigma_+ (R)$, the derived GMCF is independent of the depth
of the projection (here equal to the simulation box size), so long
as this depth is much larger than the correlation length of dark
matter.

The GMCF is estimated for two different kinds of lens galaxy
samples:
\begin{itemize}
\item
AG-sample, which contains all simulated
galaxies in some luminosity range;
\item
CG-sample, which contains only
central galaxies in some luminosity range.
\end{itemize}

To quantify the  GMCF, we fit it by simple models. Following
M2002, we first fit a GMCF with a power-law,
\begin{equation}
\Delta\Sigma_{+} =A~(R/1\mpch)^b ~h\msun \pc^{-2}\,,
\end{equation}
where $A$ and $b$ are the fitting parameters.
Motivated by the fact that the density profiles of
CDM haloes are well described by the NFW profile
(Navarro, Frenk \& White 1997), we also fit the GMCF by assuming
the halo profile to have the form:
\begin{equation}
\rho(r) = \frac{\bar{\delta}\bar{\rho}}{(r/r_{\rm s})(1+r/r_{\rm
s})^{2}}\;,
\end{equation}
where $\bar{\delta}$ is a dimensionless density amplitude, $r_{\rm
s}$ is a characteristic radius, and $\bar{\rho}$ is the mean
background density. This profile is characterized by a virial
radius $r_{200}$ (defined to be the radius within which the
average mass density is equal to 200$\bar{\rho}$), and a
dimensionless concentration parameter $c$ (or a scale radius
$r_{\rm s}=r_{200}/c$). The amplitude $\bar{\delta}$ is
related to $c$ by
\begin{equation}
\bar{\delta} = \frac{200c^{3}}{3[{\rm ln}(1+c)+c/(1+c)]}\;.
\end{equation}
The properties of the projected NFW profile are given in
Appendix A.

We fit the GMCFs by equation (\ref{ds+}), with $c$ and
$r_{200}$ as free parameters. Note that we are not fitting a
single dark halo with the NFW profile (the GMCF is a superposition
of different haloes), and so the values of $c$ and $r_{200}$ are not
related to the concentration and radius of a single halo but the
mean of many haloes. For both the power-law and NFW model,
we restrict the fits to the range of radii between
$0.05\mpch$ and $1\mpch$.

In Fig. 2 we show the results of the GMCF for both AG and CG
samples brighter than $-19.5^m$. These results are obtained using
all simulated galaxies down to this magnitude limit and do not
take into account the selection function in the observed data.
Observational selection effects will be considered in subsection
3.3. As one can see from Fig. 2, the GMCFs for the AG and CG
samples have rather different behaviour. For the CG sample, the
GMCF is strongly peaked at small radii and tends towards  zero at
large radii. In this case, the GMCF profile is well-described by
the simple models based on the projection of halo profiles. Not
surprisingly, the NFW model gives a better fit than the power-law
model. In contrast, the GMCF for the AG sample is less centrally
peaked and tends towards a constant non-zero value at large radii.
This occurs because many  satellite galaxies are located at the
outskirts of massive halos in the simulations. As demonstrated in
Appendix B, the reduction in the amplitude of the GMCF on small
scales is  due to the separation between satellite galaxies and
halo centers. The positive value at large radii is due to the
extended distribution of mass in massive haloes. Note that the
GMCF profile for the AG sample does not follow the simple models
based the projections of halo density profiles.
\begin{figure}
\centering \vskip-0.5cm
\psfig{figure=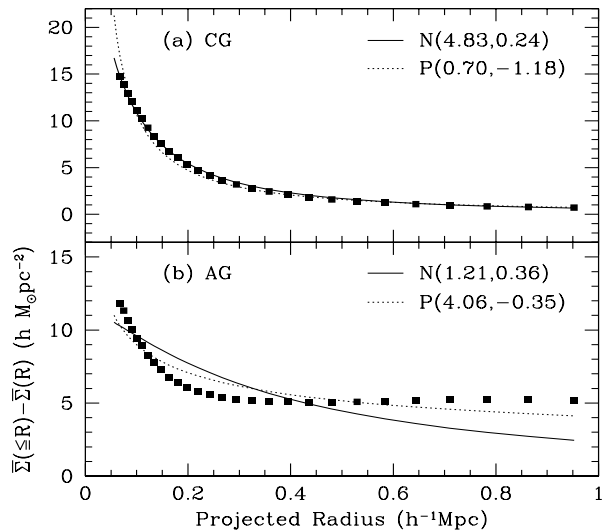,width=8.5cm,height=8.5cm,angle=0}
\vskip-0.5cm  \caption{The GMCFs for the CG (squares in the upper
panel) and AG (squares in the lower panel) samples constructed
from the GIF galaxies with $z'$-magnitudes brighter than
$-19.5^m$. The solid curve is the fit to the NFW model, with the
fitting parameters $c$ and $r_{200}$ (in $\mpch$) given in N($c,
r_{200}$). The dotted curve is the fit to the power-law model,
with the fitting parameters $A$ and $b$ given in P($A, b$).}
\end{figure}

\subsection{Dependence on galaxy properties}

Fig. 3 and 4 show how the predicted GMCFs
vary as a function of galaxy  absolute magnitude. For
bright magnitudes, most galaxies are central galaxies, and so both
AG and CG samples give similar results. For galaxies fainter than
about $-21^m$ the AG and CG samples behave differently, because
the contribution of satellite galaxies becomes important in the AG
sample (for example, over one third of the galaxies with
magnitudes in the range $-21^m$ - $-20^m$ are satellite galaxies,
see Fig. 1). There is a clear trend that more
luminous galaxies exhibit a  stronger cross-correlation with
the mass. This is expected, because
luminous galaxies reside preferentially in more massive haloes.
As we will show later, this result is reproduced quantitatively
in the observational data.

As one can see, the GMCFs are well described by the NFW model
for for central galaxies and for bright galaxies. The fit to the power-law
model is not as good, but
the difference is not big enough to
be detectable in the current SDSS data. Note that for AG samples
fainter than $-21^m$, the contamination by satellite galaxies
becomes so important that no simple model can fit the GMCFs
(see Fig. 4). These results show that the
interpretation of the GMCF is much simpler for bright central galaxies.
\begin{figure}
\centering \vskip-0.5cm
\psfig{figure=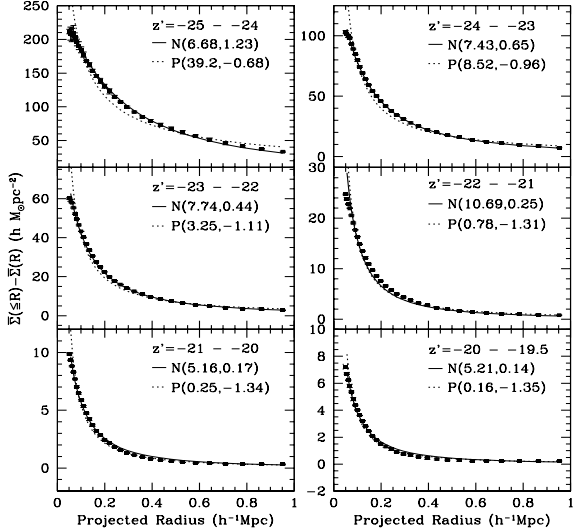,width=8.5cm,height=8.5cm,angle=0}
\vskip-0.5cm \caption{The GMCFs and the model fits for
$z$-band galaxies in different luminosity ranges (CG
samples). The best fit parameters are given as N($c,r_{200}$) for
the NFW model and as P($A,b$) for the power-law model.  The errorbars
are the scatter of the GMCFs obtained from three projections of the
simulated density field. }
\end{figure}
\begin{figure}
\centering \vskip-0.5cm
\psfig{figure=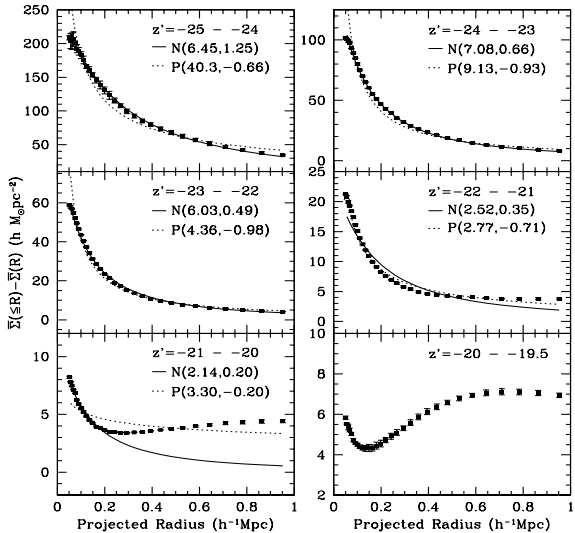,width=8.5cm,height=8.5cm,angle=0}
\vskip-0.5cm \caption {The same as Fig.3, but for AG samples. For
galaxies fainter than $-21^m$, both power-law model
and the NFW model fail to give an acceptable fit.}
\end{figure}

\begin{figure}
\centering \vskip-0.8cm
\psfig{figure=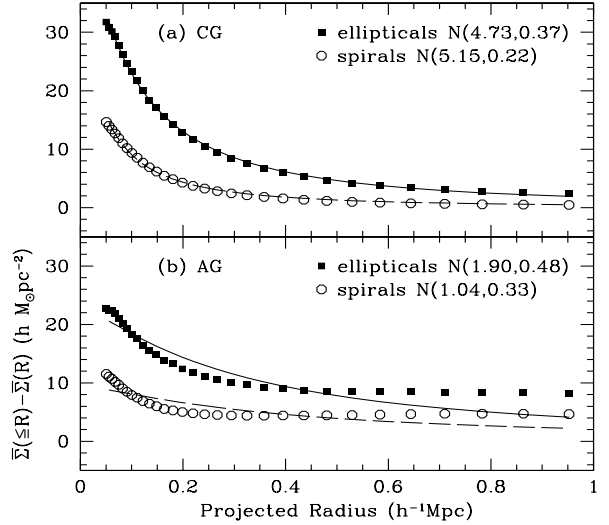,width=8.5cm,height=8.5cm,angle=0}
\vskip-0.5cm \caption{The GMCFs for elliptical (squares) and
spiral (circles) galaxies. The upper panel is for
CG samples, while the lower panel is for AG samples.
Curves show the fits to the NFW model.} \label{early_late1}
\end{figure}

In our simulation, the morphological classification of galaxies is made
according to their B-band disk-to-bulge ratios. If $M_{B, \rm
bulge}-M_{B,\rm total} < 1$ mag, the galaxy is classified as an
early-type (elliptical or S0) galaxy; otherwise it is classified
as a late-type (spiral) galaxy (see Kauffmann et al., 1999a for
details). Due to the limited simulation resolution, morphological
types are only assigned to galaxies with haloes containing more
than $\sim 100$ particles. About 7850 (13500) galaxies in the CG
(AG) sample are classified as elliptical galaxies, while about
53800 (80000) galaxies in the CG (AG) sample are classified as
spiral galaxies. In Fig. \ref{early_late1} we compare the GMCFs of
elliptical and spiral galaxies. Once  again results are
shown for all simulated galaxies without taking into
account the observational selection function. As can be seen,
the cross-corrrelation between early-type galaxies and dark matter is stronger
than for late-type galaxies.
In the simulation, these results come from the fact that early-type
galaxies are preferentially located in  massive
haloes.

\subsection {Incorporating observational selection effects}

In order to compare model predictions with observational results,
we should select galaxy samples in the same way as in the
observations. The galaxy samples analyzed in M2002 are  apparent
-- magnitude limited, and so fainter galaxies have smaller
probability of being  included. In this section, we select model
galaxies according to the observed luminosity distribution of the
lens galaxies (see figure 2 in M2002). There are some subtle effects 
associated with a magnitude - limited sample. For example, 
lens galaxies with different luminosities have different 
redshift distributions, and so their lensing strengths 
(i.e. the value of $\Sigma_{\rm crit}^{-1}$) are different.
Furthermore, the angular radius corresponding to a given 
projected physical radius is larger at a smaller distance,
and so the number of lensed sources ($N_{\rm source}$)
within a given projected physical radius is larger for closer 
galaxies. In the analyses of M2002, lens galaxxies are weighted
by $N_{\rm sources}* \Sigma_{\rm crit}^{-2}$. 
To mimic this weighting, we use the curve of 
$\Sigma_{\rm crit}^{-1}$ given in M2002 and 
assume $N_{source}\propto r^{-2}$, where $r$ is the 
angular-size distance to a lens. We calculate a mean weight 
for galaxies with a given luminosity by averaging 
$N_{\rm sources}* \Sigma_{\rm crit}^{-2}$
over the volume within the distance limit which
corresponds to the luminosity in consideration,
assuming that the distribution of such galaxies
within the volume is uniform. We then assign to each model 
galaxy a mean weight corresponding to its luminosity.
Note that the $N_{\rm source}$ used in M2002
takes into account other selection effects, such as the 
exclusion of lens galaxies around which the distributions 
of source galaxies are too asymmetric (see M2002 for details).
Such selections may change the $N_{source}\propto r^{-2}$
relation but are not taken into account in our model.  

\begin{figure}
\centering \vskip-0.8cm
\psfig{figure=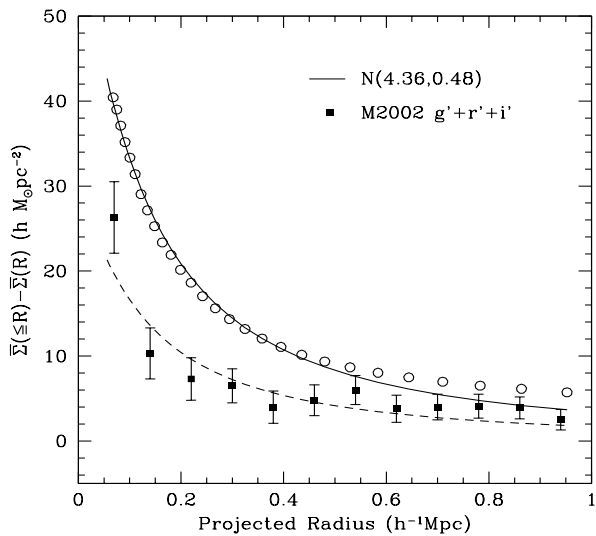,width=8.5cm,height=8.5cm,angle=0}
\vskip-0.5cm \caption{The GMCF and the model fit for the sample
constructed with the SDSS selection function. The solid line is
the fit to the NFW model, while the dashed line is half of the
model fit. For comparison, the GMCFs derived by M2002 from the
combined g'- r'- and i'-band images are shown as squares with
errorbars. Note that M2002 only used the images in these three
bands to measure the galaxy shapes. Although the predicted GMCF
has the same shape as the observation, the predicted amplitude is
too high by a factor of $\sim 2$.} \label{selected_all}
\end{figure}

Fig.\ref{selected_all} shows the results of the GMCF for a sample
of galaxies selected in  the $z$-band. It turns out that most of
the galaxies in the selected samples are luminous central galaxies
and as a result, the GMCF can now be fitted reasonably well by the
NFW model. Compared with the GMCFs derived by M2002, our predicted
GMCF has a similar shape but a higher amplitude which corresponds
to more massive halos. 

\begin{figure}
\centering \vskip-0.8cm
\psfig{figure=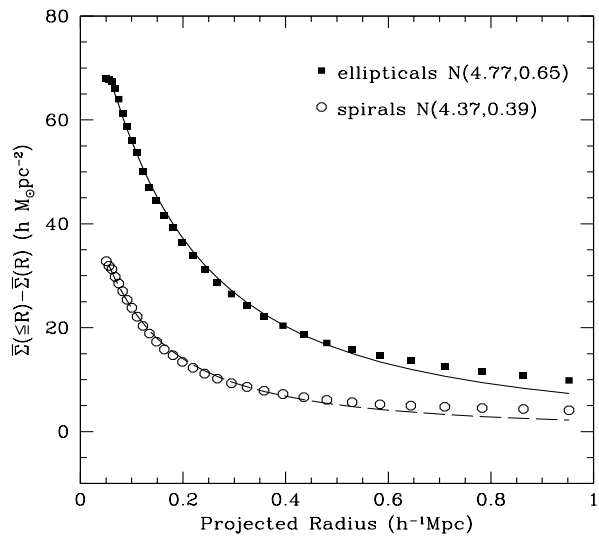,width=8.5cm,height=8.5cm,angle=0}
\vskip-0.5cm \caption{The GMCFs for elliptical (squares) and
spiral (circles) samples with the SDSS selection function. Curves
are fits to the NFW model. } \label{selected_type}
\end{figure}

When observational selection is applied to the GIF simulation,
about 3200 galaxies are classified as elliptical galaxies, while
about 7500 galaxies  are classified as spiral galaxies. The
fraction of the early-type galaxies in the simulation is lower
than that in the SDSS sample used by M2002 (${\rm E:S}\approx
1:1$). This may be due to the fact that the definitions used to
define early- and late-type galaxies are different in the GIF
simulation and in the observations. As mentioned above, galaxies
in the GIF simulation are assigned morphological types according
to their bulge/disk ratios, while galaxies used in M2002 are
classified according to their color, concentration and asymmetry.
In Fig.\ref{selected_type} we compare the GMCFs of elliptical and
spiral galaxies. Early-type galaxies have systematically higher
GMCF than late-type galaxies, consistent with the observational
results of M2002. The amplitude of GMCF is higher than the
observational results. We will discuss this in more detail in the
next section.

We have also studied the effect of environment on the GMCF.  We
count the number of neighbours with projected distances less than
$1\mpch$ from each galaxy and divide galaxies into two high- and
low-density subsamples containing equal numbers of objects. This
procedure is is similar in spirit, but somewhat simpler than the
one adopted by M2002, who defined the local density at the
position of a galaxy by the inverse of the area of the polygon
obtained from a Voronoi tesselation of the galaxy distribution on
the sky.

\begin{figure}
\centering \vskip-0.8cm
\psfig{figure=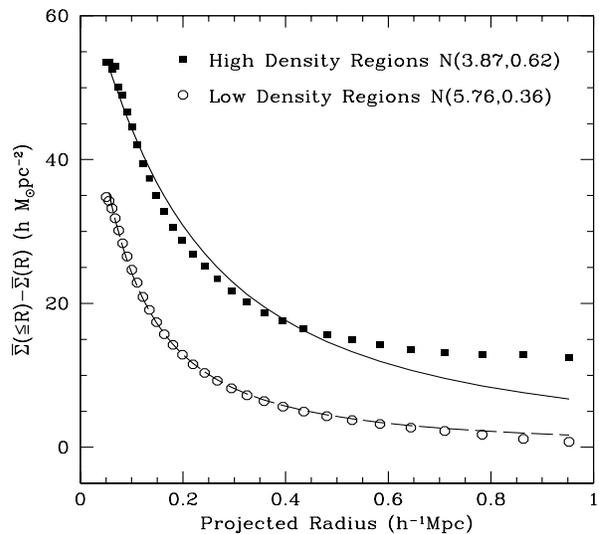,width=8.5cm,height=8.5cm,angle=0}
\vskip-0.5cm \caption{The GMCF for galaxies in high- (squares) and
low-density (circles) regions, selected with the SDSS selection
function. Curves are fits to the NFW model.} \label{selected_env}
\end{figure}

The mean $z^\prime$-band luminosities for the high and low-density
samples are 3.3 and 2.0 $\times 10^{10}{\rm h}^{-2}L_{\odot}$. The
GMCFs for these samples are shown in Fig.\ref{selected_env}.
Galaxies in low-density environments are mainly central galaxies
and as a result, their GMCF is fit very well by an NFW halo.
Galaxies in high-density environments occur in groups and
clusters. Many of them are not central galaxies-- this is why the
NFW model is a considerably worse fit.

\section {The mass-luminosity relations}

One important goal of studying galaxy-galaxy lensing is to derive
the relation between luminosities of galaxies and the masses of
their host haloes. In this section, we study whether or not the
observed GMCF can be used to infer such a mass-luminosity
relation.

\subsection{The $M_{260}$-$L$ Relation}

In their paper, M2002 attempted to estimate the halo mass within a
fixed aperture of radius $R=260\kpch$. The reason for this choice
is that the GMCF at a much larger radius is likely to be  affected
by projection effects, and at small radius, the GMCF is difficult
to measure observationally (see M2002 for detailed discussion).

In this section we follow the procedure adopted by M2002 and fit
the GMCF to a projected singular isothermal sphere (SIS) model to
obtain $M_{260}$ for the galaxy samples we have selected out of
the simulations:
\begin{equation} \label{SIS}
\Delta_+\Sigma(R)_{SIS} = \Sigma(R)_{SIS} = { \sigma_v^2 \over 2 G
} {1 \over R}\,,
\end{equation}
where $\sigma_v$ is the line-of-sight velocity dispersion. We have
also fit the GMCF to a NFW profile, which gives the parameters
$r_{200}$ and $c$ (see Subsection 2.2). The aperture mass
$M_{260}$ is then the integral of $\Delta\Sigma_{+}$ within a
radius $260\kpch$.

\begin{figure}
\centering \vskip-0.8cm
\psfig{figure=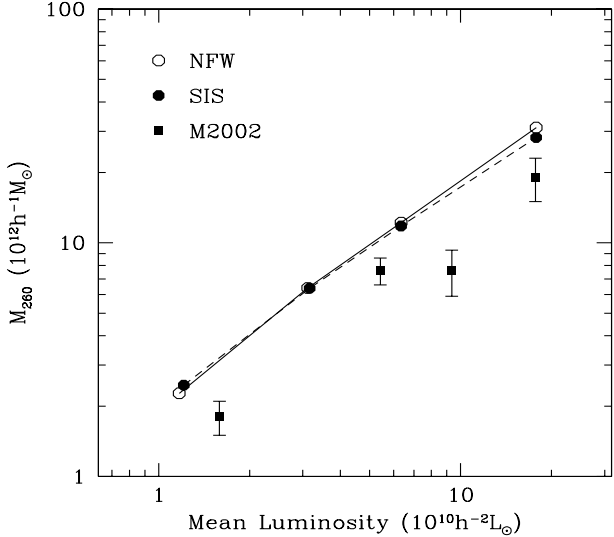,width=8.5cm,height=8.5cm,angle=0}
\vskip-0.5cm \caption{The $M_{260}$-luminosity relation for
galaxies in the GIF simulation. Open circles are based on the fit
to the NFW model, while solid circles are based on the fit to the
SIS model. The squares with errorbars are the results from M2002.}
\label{M260asL}
\end{figure}

Fig.\ref{M260asL} shows $M_{260}$ as a function of the mean
luminosity of galaxies. Both the SIS and NFW models give very
similar aperture masses. For comparison, we also display the
$M_{260}$-$L$ relation given by M2002. The predicted trend of mass
with luminosity is similar to the observation, but the predicted
$M_{260}$ is about a factor of 2 larger than the observed value.

\begin{figure}
\centering \vskip-0.8cm
\psfig{figure=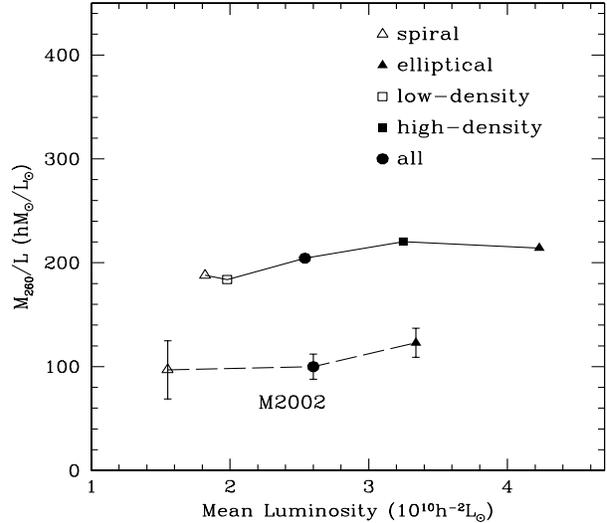,width=8.5cm,height=8.5cm,angle=0}
\vskip-0.5cm  \caption {The $M_{260}/L$ ratios for various samples
taking into account the observational selections. The mass/light
ratios are measured separately for spiral and elliptical galaxies,
galaxies in low- and high-density regions, and all galaxies. The
results of various simulation samples are connected with a solid
line. For comparison, the results of M2002 are shown by symbols
connected with a long dashed line.} \label{M260toL}
\end{figure}

In Fig.\ref{M260toL} we show the predicted $M_{260}/L$ ratios for
various simulated samples, and compare them with the measurements
of M2002. Note that the observational selection is applied to the
total sample, rather than to individual samples. This is the
reason why the mean luminosities of the morphological samples are
somewhat different from the observed values. We find that the
$M_{260}/L$ ratios are independent of galaxy luminosity,
morphological type and environment, consistent with the
observational results. The predicted amplitude of $M_{260}/L$ is
about two times the observed value, indicating that the model
over-predicts the halo mass (or under-estimates the luminosity).

As shown above, the predicted mass-to-light ratios in the
simulations are about two times that of the observed value. Should
we take this discrepancy seriously? In the GIF catalogs, model
parameters are calibrated to reproduce the observed Tully-Fisher
(TF) relation. This calibration  depends on how the observed
rotation velocities are related to the halo circular velocities
and is hence uncertain. Kauffmann et al
(1999) have shown that their model underpredict the 
number density of $L_*$ galaxies by a factor of 2, and
so a higher luminosity for a given halo mass may 
make the model prediction in better agreement with the 
observed luminosity function around $L_*$.
However, assigning much higher luminosities to dark 
haloes may lead to too high a luminosity density for the
local universe (Kauffmann et al. 1999; see also
Yang, Mo \& van den Bosch 2002). 
  
\subsection { Deriving  halo masses from the GMCFs}

\begin{figure}
\centering \vskip-0.8cm
\psfig{figure=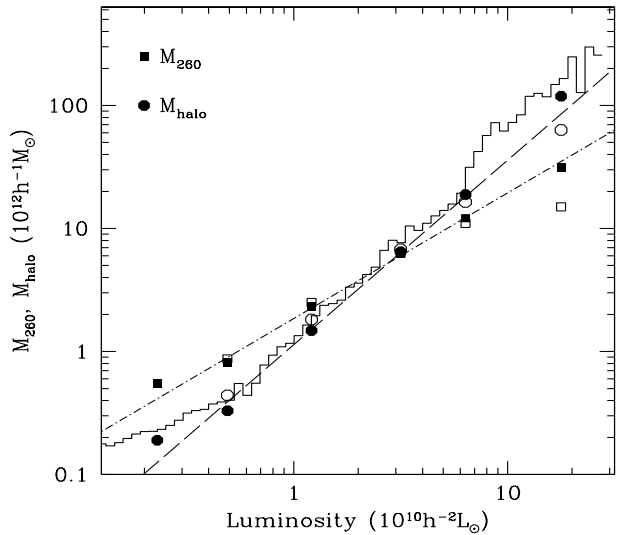,width=8.5cm,height=8.5cm,angle=0}
\vskip-0.5cm \caption{The aperture mass $M_{260}$ (squares) and
halo mass $M_{\rm {halo}}$ (circles) recovered from the GMCFs,
together with the mean halo mass directly measured from the
simulations (histogram). Solid symbols are for CG samples and open
symbols are for low-density samples taking into account of
observational selection. The two straight lines show power laws
$M\propto L^{1.5}$ and $M\propto L$, respectively. Note that over
a large luminosity range, the aperture mass $M_{260}$ in the
simulation increases linearly with $L$, while the halo mass
increases roughly as $L^{1.5}$.} \label{mhalo and m260}
\end{figure}

The relation between the  aperture masses $M_{260}$ and the mass
of a galaxy's halo is complicated. We have already highlighted one
important problem-- for ``non-central" galaxies in groups,
$M_{260}$ will have almost no relation to the mass of the
surrounding dark halo. Even if one were to impose an isolation
criterion on the sample and select only central galaxies, the
relation between $M_{260}$ and $M_{halo}$ is not trivial. For
luminous central galaxies, $M_{260}$ only measures the mass in the
inner regions of their halos, while for faint central galaxies,
$M_{260}$ measures the mass beyond the virial radius of the halo .
This is shown in Fig. \ref{mhalo and m260}. As one can see,
$M_{260}$ is lower than the mean halo mass for bright central
galaxies and is higher than the mean halo mass for faint central
galaxies. Over a large range of luminosity, the aperture mass
$M_{260}$ scales linearly with galaxy luminosity, while the halo
mass scales as $L^{1.5}$.

\begin{figure}
\centering \vskip-0.8cm
\psfig{figure=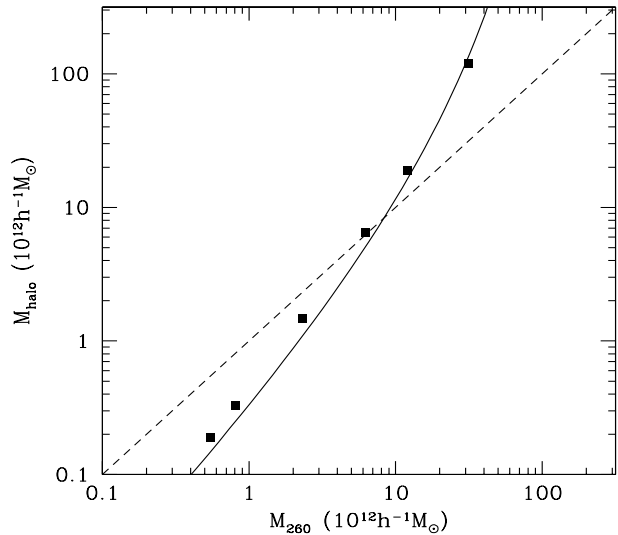,width=8.5cm,height=8.5cm,angle=0}
\vskip-0.5cm  \caption {The $M_{halo}$-$M_{260}$ relation. The
solid line is the prediction of a simple model discussed in the
text. The squares are inferred from the GMCF for CG samples.}
\label{mh260}
\end{figure}

Fig. \ref{mh260} shows the relation between $M_{\rm halo}$ and
$M_{260}$ for central galaxies in the simulation. Once again we
see that $M_{\rm halo}$ and $M_{260}$ are tightly correlated, but
the relation is not linear. For haloes more massive than
$10^{13}h^{-1}\msun$, $M_{\rm halo}$ is larger than $M_{260}$,
while for less massive haloes, $M_{\rm halo}$ is smaller than
$M_{260}$. This relation can also be obtained using a simple
analytical model. From numerical simulations, it is known that the
concentration of a halo  $c$ is correlated with $M$. We use the
$c$ -- $M$ relation obtained by Bullock et al. (2002):
\begin{equation}\label{conc}
 c(M)\approx 9\left({M\over M_*}\right)^{-0.13}\,,
\end{equation}
where $M_*$ is the characteristic non-linear mass scale at which
the $rms$ of the linear density field is equal to the critical
overdensity for spherical collapse, $\sigma (M_*)=\delta_c\sim
1.69$. The definition of the halo mass $M$ is somewhat different
from ours, but this relation changes only a little. With this
model, we can predict $M_{\rm halo}$ as a function of $M_{260}$.
The solid curve in Fig. \ref{mh260} shows this prediction.

We now ask how one can derive halo masses from the full GMCF. One
possibility is to use the results of the NFW fit to the GMCF to
obtain an average `halo mass' (denoted by $M_{\rm halo}$) for a
set of lens galaxies that have no close neighbours of comparable
brightness and are thus likely to be ``central" objects.   In Fig.
\ref{mhalo and m260} we plot the mass recovered this way as a
function of galaxy luminosity for the central galaxies in our
simulation (solid dots). A comparison with the mean halo mass
measured directly from the simulation shows that the agreement is
very good over a large range of halo masses. Thus, for central
galaxies, this method works well in recovering the mean halo mass.
The open circles in Fig. \ref{mhalo and m260} show the recovered
halo masses for the sample of galaxies in low-density environments
defined in section 3.3.  The results are very close to that for
central galaxies, except at the very bright end where the halo
mass is lower because bright galaxies in massive, rich clusters
are missing. Thus, it is also possible to obtain the halo mass of
galaxies by using the GMCFs of galaxies in low-density
environments. In the next subsection we will apply this method to
the data of M2002 to derive halo masses for the observed galaxies
as a function of luminosity.

The second method for deriving halo masses is to make use of the
relation between $M_{\rm halo}$ and $M_{260}$ obtained from
simulations (Fig. \ref{mh260}). This will be described in more
detail in the next section.

\subsection {Halo masses of the observed galaxies}

\begin{figure}
\centering \vskip-0.8cm
\psfig{figure=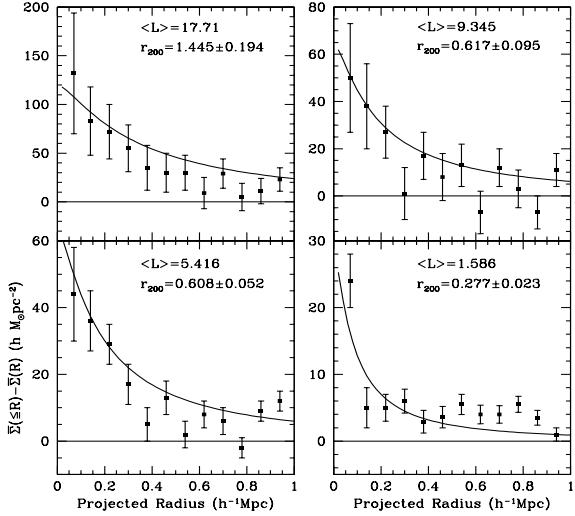,width=8.5cm,height=8.5cm,angle=0}
\vskip-0.5cm  \caption {The NFW fits to the observed $z^\prime$
band GMCF data in Fig. 15 of M2002. Here we assume that the
concentration $c$ depends on the halo mass as Eq. (\ref{conc}),
and use $r_{200}$ as the only free parameter. The fits only extend
over radii smaller than $260\kpch$, as in
M2002.}\label{GMCF_M2002}
\end{figure}

\begin{figure}
\centering \vskip-0.8cm
\psfig{figure=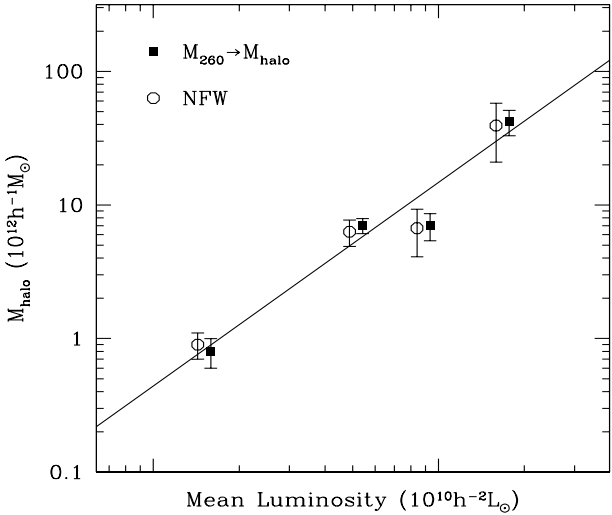,width=8.5cm,height=8.5cm,angle=0}
\vskip-0.5cm  \caption {The derived  halo masses $M_{halo}$ as a
function of luminosity. The squares show $M_{halo}$ estimated
using the relation shown in Fig.\ref{mh260}. The circles are the
halo masses  obtained from the fits in Fig. 16. Errorbars are
estimated using 1-$\sigma$ variances of the data points. To avoid
confusion, the results of the NFW fit are shifted to the left. The
solid line is a power law fit of the data points shown.}
\label{observ}
\end{figure}

We now apply our two methods of estimating halo masses to the
actual observational data. Fig.\ref{GMCF_M2002} shows the results
of fitting a NFW profile to the GMCFs of M2002 (their Fig. 15). As
in M2002, we have used only data points at projected radii smaller
than $260\kpch$. In order to obtain stable results, we have
assumed that the concentration $c$ depends on halo mass as in
equation (\ref{conc}). The open circles in Fig.\ref{observ} show
the halo mass obtained in this way as a function of galaxy
luminosity. The errorbars on the derived masses assume that the
observed errors on the  GMCF at different radii are independent
and are  Gaussian distributed. Since the observed errors at
different radii are actually correlated, the real errorbars on
$M_{\rm halo}$ should be smaller than those shown in the figure.

The other way to derive $M_{\rm halo}$ is to convert the aperture mass
$M_{260}$ into a halo mass using the relation shown in
Fig.\ref{mh260}. These results are shown as  solid squares in
Fig.\ref{observ}, and the errorbars are obtained directly from the
errorbars on the observed $M_{260}$. The halo masses derived from
these two methods are very similar.

As one can see, the halo mass increases with luminosity roughly as
a power law. A simple linear regression of all the data points
gives
\begin{equation}\label{mhaloasL}
M_{\rm halo}
\approx 2.0\times 10^{12} h^{-1}{\msun}
\left({L\over L_*}\right)^{1.5}\,,
\end{equation}
where $L_*\approx 2.56\times 10^{10} h^{-2}{\rm L}_\odot$ is the
characteristic luminosity in the $z'$-band (Blanton et al. 2001).
This scaling is similar to that obtained from the GIF
simulation (see Fig.~\ref{mhalo and m260}).
In terms of mass-to-light ratio, we can write
\begin{equation}
{M_{\rm halo}\over L}
\approx 75 h
\left({L\over L_*}\right)^{0.5}
{\msun\over {\rm L}_\odot}\,.
\end{equation}
This mass-to-light ratio is about two times smaller than the
simulation results. The origin of this discrepancy has already
been discussed in Subsection 4.1.

The power index $1.5$ we obtain for the $M_{\rm halo}$--$L$
relation is similar to that derived by Guzik \& Seljak (2002)
using an analytic model of the galaxy distribution in dark haloes.
However, our value of the mass-to-light ratio for $L_*$ galaxies
is about two times larger than theirs. One possible reason for
this discrepancy is that Guzik and Seljak assume that galaxies in
groups and clusters have the same halo masses as isolated galaxies
of the same luminosity. If galaxy halos in groups are tidally
truncated, their assumption will cause the halo masses of isolated
galaxies to be underestimated. Current high-resolution N-body 
simulations (e.g. Klypin et al. 1999; Moore et al. 1999;
Springel et al. 2001) show that subhalos of dark matter in
galaxy systems have radii typically much smaller than 
$200\kpc$, and so their contribution to the lensing 
signals on the scales we are concerned here
should not be important. 

Since the ratio  $M_{260}/L$ does not depend on morphological
type, we may apply our method  to spirals to check whether their
halo masses are consistent with those derived from the
Tully-Fisher relation (e.g. Hudson et al. 1998; Smith et al. 2001;
Wilson et al. 2001; Seljak 2002).

In the $I$-band, the observation of Giovanelli et al. (1997) gives
\begin{equation}
L_I\approx 1.0\times 10^{10}\left({V_{\rm obs}\over 150\kms}\right)^{3.1}
h^{-2} {\rm L}_{\odot I}\,,
\end{equation}
where $V_{\rm obs}$ is an observational measure of the rotation
velocity, usually taken as the maximum rotation velocity. The
typical  $z'-I$ colour of a spiral galaxy is similar to that  of
the Sun, so its luminosity in solar units will have the same value
in different bands. Thus, for an $L_*$ galaxy with $z'$-band
luminosity $2.56\times 10^{10} h^{-2}{\rm L}_\odot$, the
Tully-Fisher relation implies $V_{\rm obs}\approx 205\kms$. If we
include an internal extinction of about $0.3^m$ (see Verheijen
2001) this will increase $V_{\rm obs}$ to about $225\kms$. The
halo mass for an $L_*$ galaxies obtained from the Tully-Fisher
relation is thus
\begin{equation}
M_{\rm halo,*}\sim
4.8\times 10^{12} h^{-1} \left({V_c\over V_{\rm obs}}\right)^3\msun\,,
\end{equation}
where $V_c$ is the circular velocity of the halo. We have assumed
$\Omega_0=0.3$, and defined the halo radius such that the mean
density within it is 200 times the mean density of the Universe.
Comparing this with the lensing result given in equation
(\ref{mhaloasL}), we see we require $V_{\rm obs}/V_c\sim 1.34$.
This ratio is consistent with the predictions of disk formation
models (e.g. Mo, Mao \& White 1998; Mo \& Mao 2000). The $V_{\rm
obs}/V_c$ ratio obtained here is lower than that obtained by
Seljak (2002) based on the mass determination published in Guzik
\& Seljak (2002). This is  because our estimated halo mass is
higher than theirs.

\section{Discussion and Summary}

In this paper we have used galaxy catalogues constructed from a
high-resolution numerical simulation to investigate the properties
of the projected galaxy-mass correlation function (GMCF) measured
from weak galaxy-galaxy lensing observations. A comparison between
the predicted GMCFs with those obtained recently by McKay et al.
(2002) from the Sloan Digital Sky Survey (SDSS) data shows that
the observed dependence of the GMCF on galaxy luminosity,
morphological type and environment is well reproduced in the
model. In the simulations, the ratio between the aperture mass
within a radius of 260$\kpch$ and lens galaxy luminosity is
independent of galaxy luminosity, type and environment. The value
of $M_{260}/L$ is approximately equal to the ratio between the
halo mass (defined to be the mass within the halo virial radius)
and luminosity for $L_*$ galaxies. $M_{260}$ overestimates the
halo mass for lens galaxies with $L\ll L_*$ and underestimates the
halo mass for $L\gg L_*$.

Using the simulations, we show that  halo masses  can be recovered
by fitting the full GMCF with  realistic halo density profiles. If
we apply this method to the observational data, we find that the
halo mass scales with galaxy luminosity as $M_{\rm halo}\propto
L^{1.5}$. The mean halo mass we derive for an $L_*$ spiral galaxy
is consistent with that inferred from the observed  Tully-Fisher
relation.

We find that the behavior of the GMCF can be significantly
affected by satellite galaxies in high-density environments. The
observed GMCF can only be used  to probe the properties of
individual galaxy haloes if galaxy samples are limited to central
galaxies. We have demonstrated that an effective way of doing this
is to select bright and relatively isolated galaxies. In the
future when lensing signal can also be detected for galaxies
fainter than $L_*$ (many of which are expected to be satellite
galaxies), detailed modelling of the galaxy distribution in dark
halos will be  required to compare models and observations.

Our results suggest that the GMCF obtained from galaxy-galaxy
lensing can provide important constraints on galaxy formation models.
With the completion of the SDSS, such constraints will become an
integral part of our understanding of galaxy formation.

\section*{Acknowledgement}
We thank Simon White for useful suggestions and
the GIF group for the public release of their $N$-body
simulation data. XHY thanks the CAS-MPG exchange program 
for support.

\appendix

\section{The projection of NFW profile}

The properties of the projected NFW profile can be found in, {\it
e.g.}, Bartelmann (1996). The surface mass density can be
expressed as a function of a dimensionless radius $x\equiv
R/r_{\rm s}$:
\begin{equation}\label{sig}
\Sigma(x)=A~f(x)
\end{equation}
where
\begin{eqnarray}
f(x<1)&=&\frac{1}{x^{2}-1}\left(1-\frac{{\ln
{\frac{1+\sqrt{1-x^2}}{x}}}}{\sqrt{1-x^{2}}}\right)\nonumber\\
f(x=1)&=&\frac{1}{3}\\
f(x>1)&=&\frac{1}{x^{2}-1}\left(1-\frac{{\rm
atan}\sqrt{x^2-1}}{\sqrt{x^{2}-1}}\right)\nonumber
\end{eqnarray}
and
\begin{equation}
A=2~r_{s}~\bar{\delta}~\bar{\rho_{c}}\;.
\end{equation}
The mean surface mass density within radius $x$ is:
\begin{equation}
\Sigma(\leq x)=A~g(x)
\end{equation}
where
\begin{eqnarray}
g(x<1)&=&\frac{2}{x^{2}}\left(\ln\frac{x}{2}+ \frac
{{\ln{\frac{1+\sqrt{1-x^2}}{x}}}}{\sqrt{1-x^{2}}}\right)\nonumber\\
g(x=1)&=&2+2\ln\frac{1}{2}\\
g(x>1)&=&\frac{2}{x^{2}}\left(\ln\frac{x}{2}+ \frac{{\rm
atan}\sqrt{x^2-1}}{\sqrt{x^{2}-1}}\right)\nonumber
\end{eqnarray}
The mass density contrast is simply
\begin{equation}\label{ds+}
\Delta\Sigma_+(x)=A~[g(x)~-~f(x)]\,.
\end{equation}

\section{Contribution by satellite galaxies}

If the lens galaxy is not located at the center of its halo, the
average surface mass density and the GMCF will depend on an extra
parameter $R_{c}$, the distance between the galaxy and the center
of the halo. The average surface mass density in this case is
\begin{equation}
\Sigma^\prime(r) = \frac {1}{2\pi} \int_{0}^{2\pi} \Sigma(r') {\rm
d}\theta
\end{equation}
where $\Sigma(r')$ is the NFW surface mass density and
$r'=\sqrt{R_c^2+r^2+2R_cr\cos{\theta}}$.

In Fig. B1 we show the surface mass density (upper panel) and the
density contrast (lower panel) as a function of radius $R$ for
various choices of $R_c$. The parameters are set to be
$r_{200}=1.2\mpch$, $c=6.0$, and $R_c$ changes from $0$ to
$0.4\mpch$. The figure shows that the surface mass density has a
peak near $R_c$, while the density contrast reaches a minimum near
$R_c$ and then increases rapidly with increasing radius. Clearly,
the behavior of the GMCF for satellite galaxy is very different
from that for central galaxy (corresponding to $R_c=0$).
The complex behavior of the GMCF of
satellite galaxies complicates the interpretation of the observed
GMCF, and so it is important to separate central and satellite
galaxies in the galaxy-galaxy lensing study. As we have
demonstrated above, selections of central galaxies can be made
quite effectively by using bright and relatively isolated
galaxies.
\begin{figure}
\centering \vskip-0.8cm
\psfig{figure=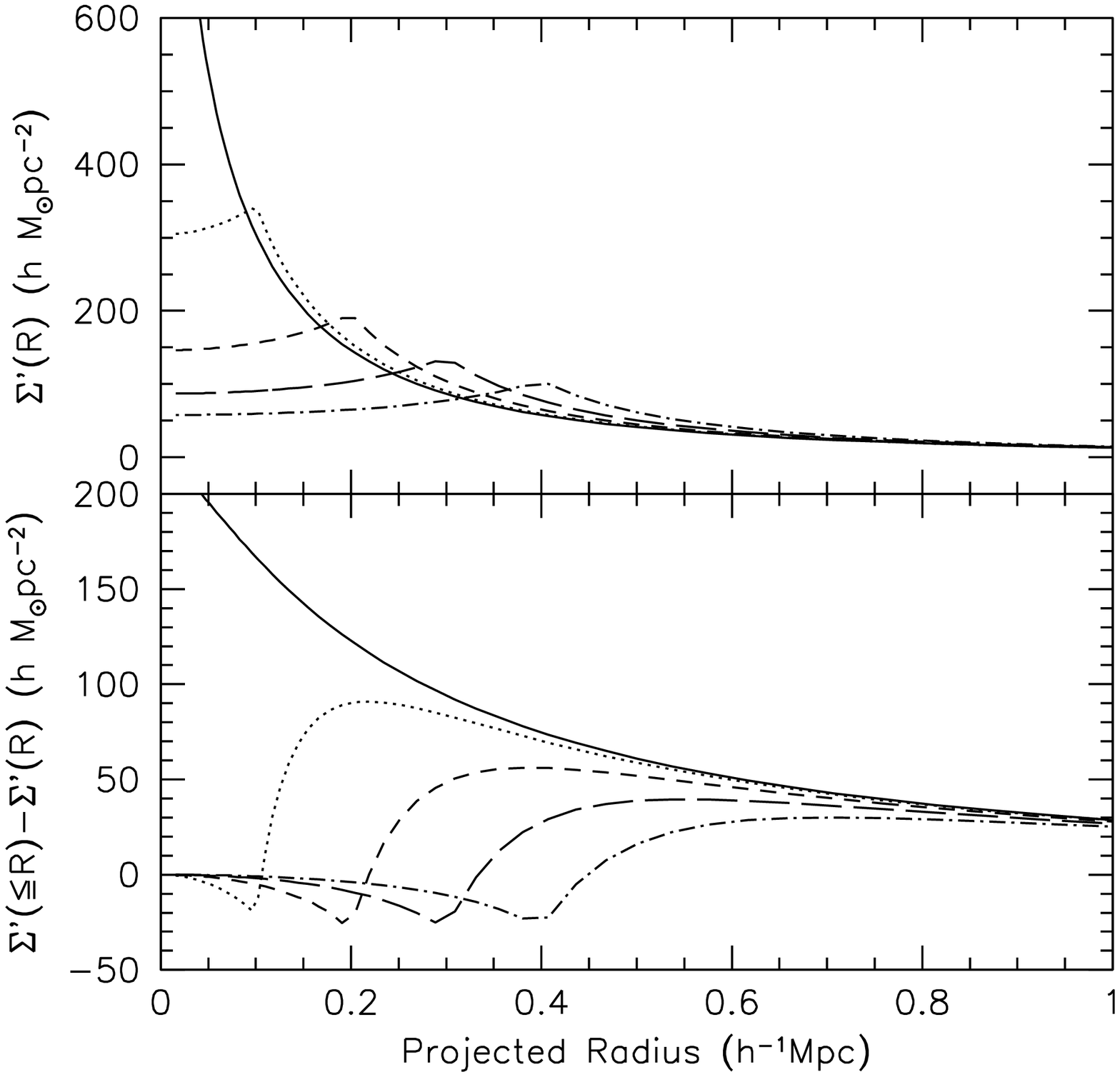,width=8.5cm,height=8.5cm,angle=0}
\vskip-0.5cm \caption{The surface density (upper panel) and
density contrast (lower panel) around satellite galaxies as a
function of radius $R$. The results are shown for a single NFW
halo with $r_{200}=1.2\mpch$ and $c=6.0$. Different curves
correspond to different distances between the satellite and the
halo center: $R_c=0$ (solid); 0.1 (dotted); 0.2 (dashed); 0.3
(long dashed); 0.4 (dot-dashed) (all in units of $\mpch$).}
\end{figure}

\end{document}